\documentclass[pdflatex,iicol,sn-mathphys-num]{sn-jnl}

\usepackage{natbib}
\usepackage{graphicx}%
\usepackage{multirow}%
\usepackage{amsmath,amssymb,amsfonts}%
\usepackage{amsthm}%
\usepackage{mathrsfs}%
\usepackage[title]{appendix}%
\usepackage{xcolor}%
\usepackage{textcomp}%
\usepackage{manyfoot}%
\usepackage{booktabs}%
\usepackage{algorithm}%
\usepackage{algorithmicx}%
\usepackage{algpseudocode}%
\usepackage{listings}%
\usepackage{subcaption}
\usepackage{url}
\usepackage{mathrsfs}
\usepackage{tikz}

\theoremstyle{thmstyleone}%
\theoremstyle{thmstyletwo}%

\theoremstyle{thmstylethree}%

\begin{document}

\newcommand\submittedtext{%
  \footnotesize Preprint. Manuscript submitted to \emph{ 	
International Journal on Software and Systems Modeling}
  (Springer). Copyright may be transferred without notice, after which this version may no longer be accessible.}

\newcommand\submittednotice{%
\begin{tikzpicture}[remember picture,overlay]
\node[anchor=south,yshift=10pt] at (current page.south) {\fbox{\parbox{\dimexpr0.65\textwidth-\fboxsep-\fboxrule\relax}{\submittedtext}}};
\end{tikzpicture}%
}

\title[Taxonomic Trace Link]{Taxonomic Trace Links: Rethinking Traceability and its Benefits}

\author[1]{\fnm{Waleed} \sur{Abdeen}}\email{waleed.abdeen@bth.se}

\author[1]{\fnm{Michael} \sur{Unterkalmsteiner}}\email{michael.unterkalmsteiner@bth.se}

\author[2]{\fnm{Alexandros} \sur{Chirtoglou}}\email{alexandros.chirtoglou@hochtief.de}

\author[2]{\fnm{Christoph Paul} \sur{Schimanski}}\email{christoph.schimanski@hochtief.de}

\author[2]{\fnm{Heja} \sur{Goli}}\email{heja.goli@hochtieve.de}

\author[1]{\fnm{Krzysztof} \sur{Wnuk}}\email{krzysztof.wnuk@bth.se}
 
\affil[1]{\orgdiv{Software Engineering}, \orgname{Blekinge Institute of Technology}, \orgaddress{\city{Karlskrona}, \country{Sweden}}}

\affil[2]{\orgdiv{ViCon}, \orgname{HOCHTIEF GmbH}, \orgaddress{\city{Essen}, \country{Germany}}}

\abstract{Traceability greatly supports knowledge-intensive tasks, e.g., coverage check and impact analysis. Despite its clear benefits, the \emph{practical} implementation of traceability poses significant challenges, leading to a reduced focus on the creation and maintenance of trace links. We propose a new approach --- Taxonomic Trace Links (TTL) --- which rethinks traceability and its benefits. With TTL, trace links are created indirectly through a domain-specific taxonomy, a simplified version of a domain model. TTL has the potential to address key traceability challenges, such as the granularity of trace links, the lack of a common data structure among software development artifacts, and unclear responsibility for traceability. We explain how TTL addresses these challenges and perform an initial validation with practitioners. We identified six challenges associated with TTL implementation that need to be addressed. Finally, we propose a research roadmap to further develop and evaluate the technical solution of TTL. TTL appears to be particularly feasible in practice where a domain taxonomy is already established.}


\maketitle
\submittednotice

\section{Introduction}\label{sec:intro}

Traceability is the ability to relate data that is stored within artifacts to each other and to use these relationships to conduct analyses, such as coverage and impact analysis, that are important for systems and software engineering~\cite{gotel2012traceability}. Trace links between artifacts increase the value and usefulness of the information they record. The most common usage scenarios for trace links in requirements traceability are finding the origin and rationale of requirements, tracking requirement implementation state, analyzing requirements coverage in source code, and developing test cases based on requirements~\cite{bouillon_survey_2013}. The existence of trace links improves the software developers' efficiency and effectiveness. 
Mäder et al.~\cite{mader_developers_2015} performed an experiment involving 71 subjects with various levels of experience concluding that developers performed 24\% better on maintenance tasks and produced 50\% more correct solutions when traceability is introduced into the development process.

Despite clear benefits, the practical implementation of traceability poses significant challenges, as highlighted by various studies conducted by Fucci et al.~\cite{fucci_when_2022}, Maro et al.~\cite{maro_tracimo_2022}, Ruiz et al.~\cite{ruiz_why_2023}, and Mucha et al.~\cite{mucha_systematic_2024}. Among these challenges, three stand out that we deem to be addressed poorly by traditional trace links that connect source and target artifacts directly.

The first challenge is the granularity of traces~\cite{wohlrab_collaborative_2016,maro_tracimo_2022} as it is difficult to identify the desired level of granularity for trace links. Therefore, a trade-off must be made between the usefulness of the links and the effort required to maintain them. Additionally, artifacts created during software development have varying levels of abstraction~\cite{charalampidou_empirical_2021}, which poses a challenge for identifying direct trace links. The second challenge is the lack of a common artifact structure and tools~\cite{fucci_when_2022,mucha_systematic_2024}, which in large complex systems results in scattered information. When tools lack interoperability and the document structures vary, direct trace links are ineffective. The third challenge is unclear responsibility for establishing traceability~\cite{fucci_when_2022,ruiz_why_2023}. Creating direct links requires knowledge about both source (e.g. requirement) and destination artifacts (e.g. source code), making it often unclear who is responsible for the links. Furthermore, the creator of the trace link might not be the user of the link, causing traceability to be seen as a burden to the creators.

The trace link creation process involves trace capture and trace recovery~\cite{gotel2012traceability}. In trace capture, links are created concurrently with the artifacts that are associated with each other. This requires that both artifacts are created simultaneously. In trace recovery, existing artifacts are analyzed to identify associations between them. Trace capture has the advantage that it is easier to validate trace link fidelity while the artifacts are created, involving the creators of the artifacts, as opposed to recovery where trace links are typically not recovered by the artifact creators. Trace recovery has the advantage that trace links can be created on demand and do not need any upfront investment. A common disadvantage of trace capture and recovery is that the creator of trace links is seldom also the user/beneficiary (the role that benefits from and participates in the usage scenarios for trace links~\cite{arkley_overcoming_2005,wohlrab_collaborative_2016}). This conundrum motivated us to rethink how and, in particular, \emph{when} trace links are created and how we can design a trace process that is beneficial for the creators of trace links.

In a previous study~\cite{unterkalmsteiner_early_2020}, we proposed using an auxiliary artifact to create trace links between the source and target artifact. The key difference to traditional trace links is that artifacts are not associated directly with each other but through an auxiliary artifact, introducing a level of indirection that provides practical benefits in traceability implementation. We call this approach  \emph{taxonomic trace links} (TTL). We argue that the idea has the potential to address the previously mentioned challenges: granularity of traces~\cite{wohlrab_collaborative_2016,maro_tracimo_2022} (see Section~\ref{sec:abstraction}), lack of structure of the traced artifacts~\cite{fucci_when_2022,mucha_systematic_2024} (see Section~\ref{sec:structure}), unclear responsible for traceability~\cite{fucci_when_2022,ruiz_why_2023} (see Section~\ref{sec:time}).

In this paper, we discuss how TTL can address the previously mentioned challenges. Moreover, we report on a validation study~\cite{wieringa2014design} that we conducted with domain experts to explore the feasibility of TTL. The idea of TTL is not bound to any particular type of artifact. However, we rely on the ability to analyze textual information in artifacts. Thus, while we are investigating the feasibility of TTL in a systems engineering project, we expect that the idea transfers to software projects due to their reliance on using natural language for documenting requirements~\cite{wagner2019status}. Finally, we present a research roadmap for the practical implementation and extended validation of TTL.

The contributions of this paper can be summarized as follows:

\begin{enumerate}
\item A detailed explanation of the TTL idea aiming to enhance traceability between requirements and system artifacts
\item Presents practical insights into the application of TTL, highlighting challenges and lessons learned in from using TTL to trace requirements artifacts to other system artifacts.
\item A research roadmap for the development and evaluation of TTL
\end{enumerate}

The remainder of this paper is organized as follows. In Section~\ref{sec:ttl}, we present the idea of TTL. We explain how TTL can address traceability challenges in Section~\ref{sec:challenges}. Then, we present traceability scenarios where TTL could be beneficial in Section~\ref{sec:scenarios}. We present the validation study of TTL in Section~\ref{sec:validation}. Section~\ref{sec:roadmap} presents the lessons learned and research roadmap for developing and validating TTL in practice. We conclude the paper in Section~\ref{sec:conclusion}.

\section{Taxonomic Trace Links}\label{sec:ttl}

\begin{figure*}[htb]
\centering
\begin{subfigure}{.38\textwidth}
  \centering
  \includegraphics[width=1\linewidth]{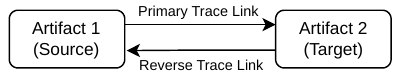}
  \caption{Traditional trace link (adapted from~\cite{gotel2012traceability})}
  \label{fig:traditional}
\end{subfigure}
\begin{subfigure}{.46\textwidth}
  \centering
  \includegraphics[width=1\linewidth]{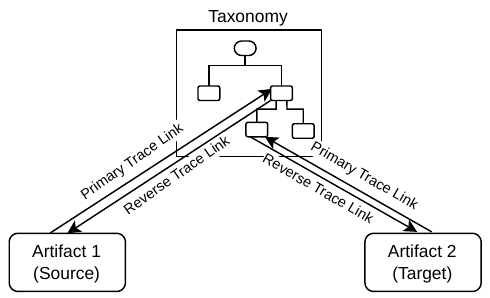}
  \caption{Taxonomic trace link}
  \label{fig:taxonomic}
\end{subfigure}
\caption{Traditional vs. taxonomic trace links}
\label{fig:tvst}
\end{figure*}

A \emph{trace} can be defined as a triplet consisting of a source and target artifact, and a primary and a reverse link associating the two artifacts~\cite{gotel2012traceability}. Figure~\ref{fig:traditional} illustrates the fundamental concept of direct trace links. The principle idea of TTL is to introduce indirection between a source and target artifact through a taxonomy. We argued~\cite{unterkalmsteiner_early_2020} that any knowledge organization structure, such as an ontology or controlled vocabularies, can be used instead of a taxonomy. To simplify communication, we coined the idea ``taxonomic trace links". Figure~\ref{fig:taxonomic} illustrates the principle of indirection introduced by links from source and target artifacts to an element in the taxonomy. 

\subsection{Domain-specific Taxonomies}

A \emph{taxonomy} is a set of nodes (classes) arranged in a hierarchy. In TTL, the taxonomy represents one (or more) dimension(s) of the problem domain, with the classes representing domain concepts. Each class has a title and could be accompanied by a description and synonyms. Tracing is performed by classifying artifacts using classes from the taxonomy, i.e., tracing is based on the semantics used in the traced artifacts.

\begin{figure}[H]
    \centering
    \includegraphics[width=1\linewidth]{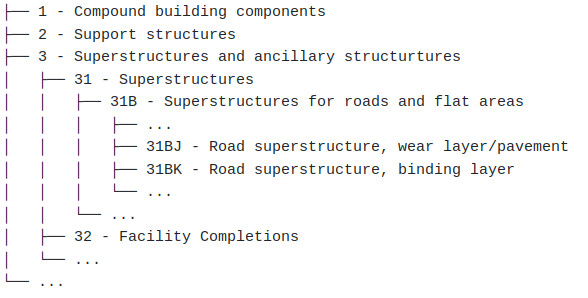}
    \caption{A Part of an Infrastructure Domain Taxonomy}
    \label{fig:domain-taxonomy}
\end{figure}

\begin{figure*}
    \centering
    \includegraphics[width=1\textwidth]{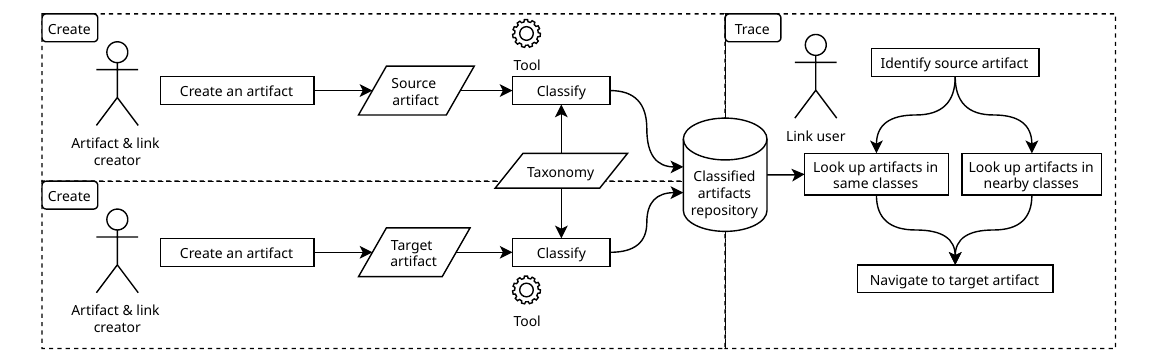}
    \caption{TTL creation and usage}
    \label{fig:ttl_create_and_trace}
\end{figure*}

An example of a domain-specific taxonomy is SB11, of which we present a subset in Figure~\ref{fig:domain-taxonomy}. The taxonomy is used in the construction industry to classify objects in models and digital twins. The elements in the taxonomy represent entities within the domain. A high-quality taxonomy should have \emph{is-a} relationships between its elements, for example, 31B: Superstructure for roads and flat areas \emph{is a} type of 31: Superstructures, which in turn \emph{is a} type of 3: Superstructures and ancillary structures. Sibling elements should not overlap; for instance, 2: Support structures does not include 1: Compound building components.

A domain-specific taxonomy can be seen as a simplified version of a domain model --- a conceptual model that captures domain knowledge in an abstract form~\cite{evans2004domain}. A domain model is typically more extensive than a taxonomy, which is a hierarchical structure of domain concepts. An ontology, on the other hand, captures even more domain knowledge by representing different types of relationships between elements. While this study focuses on using taxonomies, the approach can, in principle, use any structure that organizes knowledge systematically, provided that the additional types of relationships they capture are taken into account.

\subsection{Create and Trace}\label{sec:createandtrace}

The process of creating trace links and tracing artifacts is illustrated in Figure~\ref{fig:ttl_create_and_trace}. Depending on the domain, the taxonomy used for classification may already exist or may need to be created. Initially, as part of a software development activity, the creator generates a source artifact, e.g., a requirement. Then, the creator classifies the source artifact using a tool that predicts classes from a domain-specific taxonomy. The classified artifact is stored in a repository. By performing this classification, a trace link between the source artifact (a requirement) and the taxonomy is created. Tracing the source artifact to a target artifact requires another actor to perform the same process (potentially asynchronously) on the target artifact, e.g., a test case. The process can be repeated with different actors on different types of artifacts (e.g., classes from the source code) to trace them to the artifacts that were previously associated with the taxonomy.

To trace artifacts using TTL, the link user, who could be the creator or another actor, identifies the source artifact they want to trace. The user finds the targets by filtering artifacts using specific criteria in the repository. The filter criterion could be class equality, i.e. both artifacts are associated with the same class from the taxonomy. Other filter criteria could be that the target artifact is associated with a parent, child, sibling, or within some specified neighborhood of the class associated with the source artifact.

\subsection{Maintenance of trace links}\label{sec:maintain}

The maintenance effort required for indirect trace links is comparable to that of direct trace links, even though there may be up to twice as many links with indirect traces. For example, when tracing test cases to requirements, using direct links, we create the links: \textit{test cases → requirements}, while indirect links (TTL) require the creation of the links: \textit{test cases → taxonomy and taxonomy → requirements}. If a new test case is added ($TC_{x}$) with direct links, we need to create a trace link from $TC_{x}$ to the corresponding requirement, whereas, with indirect links, we need to create trace link(s) from $TC_{x}$ to the corresponding taxonomy class(es). The number of links required may vary depending on the domain entities referenced in $TC_{x}$.

However, when tracing more than two types of artifacts to each other, indirect trace link maintenance requires less effort than direct trace links. For example, assume we have a trace link from a requirement to two classes and two test cases. Further, assume the original requirement is split into two requirements to improve its quality (atomicity). The source code and test case links are now inaccurate. Maintaining direct trace links requires the trace maintainer to delete and add four links. On the other hand, maintaining indirect trace links in this scenario means that the trace maintainer deletes one and adds two links from the requirements to the taxonomy. With TTL, when a change occurs to an artifact, only the trace link from the artifact to the taxonomy must be updated rather than links to multiple artifacts.

\section{Traceability Challenges and Solutions}~\label{sec:challenges}

We argue that TTL has the potential to address three traceability challenges faced by practitioners, namely, \emph{granularity of traces}, \emph{lack of common data structure}, and \emph{unclear traceability responsibility}. If these challenges remain unaddressed, they could cause traceability to go awry~\cite{fucci_when_2022}. However, to what extent does indirection address these challenges? We argued in our previous work~\cite{unterkalmsteiner_tt-recs_2020} that TTL improves establishing and using trace links along three dimensions (abstraction, structure, and time). In this section, we argue for these benefits by discussing them in the context of the three named challenges.

\subsection{Granularity of Traces}\label{sec:abstraction}
The granularity of traces presents a challenge for software engineers~\cite{wohlrab_collaborative_2016,maro_tracimo_2022}. Domain concepts are often traced between artifacts on different abstraction levels~\cite{charalampidou_empirical_2021}. Borg et al.~\cite{borg_recovering_2014} studied information retrieval-based trace recovery approaches. They found that 63 out of 101 identified instances focus on artifacts on different abstraction levels (for example, code and requirements, requirements and test, or code and manuals). More recently, Charalampidou et al.~\cite{charalampidou_empirical_2021} surveyed empirical studies on software traceability and found that 122 out of 133 identified instances trace artifacts on varying abstraction levels belonging to different development activities. 
For example, a requirement specifying a customer's need might be addressed by multiple functions in the source code. When creating trace links between requirements and source code, stakeholders need to decide on tracing requirements to functions or to the parent classes of these functions. The requirements team would benefit from a high-level tracing of requirements (to classes) to identify high-risk requirements. On the other hand, the development team would benefit from more granular traces (to functions) to help change (e.g., refactoring) the source code. Eventually, a trade-off must be made between the usefulness of the links and the effort required to maintain them. Direct trace links do not support creating traces on different abstraction levels unless multiple links are created on different abstraction levels, increasing thereby the total number of traces to maintain.

The benefit of \emph{abstraction} in TTL is rooted in the four aspects of engineering information. First, information lies on a problem-solution spectrum, with requirements typically describing a non-technical problem addressed by a technical solution. Second, the problem description can describe a general concern, while a solution typically contains details that are necessary to be effective. Third, the representation of a problem is often informal, while the solution requires formalism to be repeatable. Fourth, engineering information is produced and consumed by different stakeholders with complementing skill sets for different purposes. To conclude, while complex system design and development necessitates abstraction levels of engineering information, it also leads to the need for effective means to establish traceability between those abstraction levels.

We conjecture that TTL can address the challenge of granularity of traces since the taxonomy provides inherent support for abstraction. For example, a requirement can be traced to a general concept, while the implementation addressing that requirement can be traced to a more specific concept within that branch of the taxonomy. Consequently, specifying the granularity of traces for each set of traces between two types of artifacts is not needed; instead, the granularity is determined by the taxonomy used for tracing. Furthermore, the taxonomy can be a common denominator that eases stakeholder communication.  

\subsection{Lack of a Common Data Structure}\label{sec:structure}

The lack of a common data structure of artifacts and tools~\cite{fucci_when_2022,mucha_systematic_2024} in large complex systems results in scattered information about traced artifacts across different tools with unclear structures. Engineering information about a particular development activity is often stored in dedicated systems, requiring explicit support for interoperability, such as trace link creation and use. While it is possible to use solutions that support the complete development life-cycle, it is much more common in an organization to use systems from different providers. It is even more likely to encounter interoperability issues in an outsourcing scenario or when different organizations collaborate on a product. Portillo-Rodríguez et al.~\cite{PORTILLORODRIGUEZ2012663} have compiled a list of over 130 tools that are used in distributed software development teams for coordination and communication purposes. For knowledge management tools, they highlight the importance of not only the standardization of data exchange formats but also the advancement of data integration between software engineering tools. This is particularly important when collecting and analyzing large amounts of quantitative data on both software development processes and product use~\cite{kim2017data}. The simplicity of implementation of direct links does not contribute to interoperability between artifacts and tools. A common way to achieve data integration, and thereby interoperability, is using ontologies~\cite{NEIVA2016137}.

We conjecture that TTL can reduce the structure gap by providing a common interface against which tools can be developed without knowing in advance which specific systems need to interoperate. Task-specific information management systems (for requirements, design documents, code, and test cases) can be adapted to support traces to taxonomies instead of supporting trace links to myriads of different systems.

\subsection{Unclear Responsibility}\label{sec:time}

Unclear responsibility~\cite{fucci_when_2022,ruiz_why_2023} is another challenge that could cause traceability to go awry. Creating direct trace links requires knowledge about both source and target artifacts, making it less clear who is responsible  for creating these links. Moreover, typically there is no dedicated role responsible for traceability in software projects~\cite{ruiz_why_2023}, but rather, it is common that link creation is delegated to teams other than the users of the links, causing traceability to be seen as a burden to the creators~\cite{arkley_overcoming_2005,wohlrab_collaborative_2016}. 

Artifacts that need to be traced to each other are typically not created simultaneously, but there is a time lag between the creation of the source and the target artifact. Therefore, we advocate for creating trace links as early as possible and not recovering them, which is generally costly and inaccurate. However, creating trace links early in the development lifecycle is either impossible (because the artifacts do not exist yet), or there is no incentive to create the links for the engineers as they will not use them. 
While the time gap can be reduced by certain practices such as cross-role collaboration and incremental software engineering~\cite{bjarnason2016theory}, it would be useful for trace link creation if this gap could be bridged completely. That would eliminate the need for trace link recovery at a later stage. Furthermore, traces to a taxonomy would allow for analyses that are useful for the creators of the trace links, increasing the incentives to invest the effort in traceability~\cite{arkley_overcoming_2005}.

We conjecture that TTL can address the unclear responsibility challenge, benefiting and incentivizing trace link creators. Splitting the trace link into two parts, each connected between the traced artifact and the taxonomy, clarifies who is responsible for which part of the link. The early introduction of trace links can support engineering decisions. For example, requirements specifications could be analyzed early w.r.t. their completeness and correctness~\cite{dzung_improvement_2009, kof_ontology_2010, moser_requirements_2011}, providing added value to requirements engineers. Later in the project lifecycle, when the solution is implemented and traced to the taxonomy, engineers could benefit from the existing traces when employing requirements-based verification.

\section{Traceability Scenarios}\label{sec:scenarios}

Traditional direct trace links have been widely used for multiple purposes. However, there are scenarios, e.g., software compliance and change impact analysis, where direct links tend to be too expensive to maintain or are unreliable due to domain complexity, resulting in trace links being overlooked. We argue that TTL could be a good solution in these scenarios.

\subsection{Software and System Compliance}

Software and system compliance is a necessity in regulated and safety-critical domains (e.g., medical and finance)~\cite{mubarkoot_software_2023}, where the compliance of software against specific sources, e.g., GDPR regulations, is checked. Software compliance becomes challenging when the compliance sources are multi-faced (e,g., rules, regulations, policies, and best practices)~\cite{moyon_security_2020,mubarkoot_software_2023} since the requirements engineers need to ensure that the requirements cover all the compliance sources. The same applies to infrastructure projects where local regulations, and global regulation (e.g., EU)  apply to any project which  must adhere to them. Furthermore, legal and technical experts may encounter a communication gap due to using different terminologies. This gap needs to be addressed to ensure that the software aligns with the applicable compliance sources~\cite{moyon_security_2020}. 

Using direct trace links to connect these compliance sources with software development artifacts requires proper interpretation of these sources and understanding of the terminology used in these documents. For example, in the case of regulation tracing to requirements, both law and domain expertise are required to create and maintain these links, which could be challenging due to the communication gap~\cite{mubarkoot_software_2023}.

Using TTL could bridge this gap between the technical and compliance experts by aligning these sources to a common domain taxonomy which could reduce misinterpretation caused by using different vocabulary. Moreover, connecting software development artifacts with different compliance sources becomes easier to attain since associating new compliance sources requires classification of these artifacts using the same taxonomy. This opens the opportunity to enable continuous compliance~\cite{santilli2023continuous} by associating compliance sources to different development stages, e.g., design, development, and testing.

\subsection{Change Impact Analysis}

Change management is particularly important for evolving software with continuously changing requirements~\cite{mas_traceability-why_2012}, where the artifacts produced during software development (e.g., source code, design diagrams, and test cases) are kept up to date. The lack of proper change management could lead to outdated artifacts, which consequently risk the quality of the software. Thus, a change impact analysis, enabled by reliable trace links between the produced artifacts, becomes necessary.

Using direct trace links requires that all produced artifacts be connected together to answer the query: \emph{for a given element X in artifact A, which parts of which artifacts need to be changed to introduce the change and not break existing functionalities?}. However, these links need to be complete, reliable, and accessible to the engineer performing the change impact analysis, which is not always the case~\cite{mas_traceability-why_2012,fucci_when_2022}. Trace links may be difficult to access when they are hidden in a specific tool, e.g., in a commit message~\cite{fucci_when_2022}; in this case, performing change impact analysis from the requirement engineering perspective becomes challenging as engineers need to find these commit messages, parse them and figure out the trace links. Furthermore, the change impact on direct trace links could be challenging when we trace different types of artifacts (requirements, source code, test cases). If a requirement is changed, all trace links from the changed requirement to the related source code and test cases need to be updated, which increases the probability of the trace links becoming outdated~\cite{mas_traceability-why_2012,fucci_when_2022}. Consequently, the change impact analysis results become unreliable.

Using TTL could be a better alternative, as the links are created to a centralized and fixed point (taxonomy). Accessing these links could be simpler with tool support~\cite{demuth_designspace_2015,nesic_building_2020}, where all trace links are stored in one place, making the analysis task less time-consuming. Furthermore, the impact of the change on taxonomic trace links could be less when we trace different types of artifacts (requirements, source code, test cases). If a requirement is changed, only the trace links from the changed requirement to the taxonomy need to be updated. With the help of automation, taxonomic trace links maintenance could be more reliable as the knowledge required by the creator of the links mainly concerns the created artifact and the taxonomy, resulting in more complete links.

\section{Validation Study of Creating and Using Taxonomic Trace Links}\label{sec:validation}

\begin{table*}[htb]
    \centering
    
    \caption{Participants in the validation study workshops with years of experience in the domain and in research}
        \begin{tabular}{ccccc}
        \toprule
            \textbf{Author} &
            \textbf{Affiliation} & 
            \textbf{Role} & 
            \textbf{Domain} &
            \textbf{Research}\\
        \midrule
            1 & BTH & Researcher & none & 3 \\
            2 & BTH & Researcher & none & 13 \\
            3 & HOCHTIEF & BIM and information manager & 7 & 1\\
            4 & HOCHTIEF & BIM manager and consultant & 5 & 4 \\
            5 & HOCHTIEF & Civil engineer & $\prec$ 1 & $\prec$ 1 \\
            6 & BTH & Researcher & none  & 14 \\
            
        \bottomrule
        \end{tabular}
    \label{tab:workshops_participants}
\end{table*}

The purpose of the validation study~\cite{wieringa2014design} is to better understand the practical challenges that may arise when implementing the idea of TTL. By involving practitioners, we also aim to validate the idea in a realistic setting and collect data that can help us further develop the approach. We conducted therefore a field experiment~\cite{wieringa_design_2009} involving two partners. Trafikverket, the Swedish authority responsible for planning and maintaining infrastructure such as roads, railways, and airports, provided the study material, originating from an ongoing railway project (Eastlink\footnote{\url{https://www.trafikverket.se/en/startpage/projects/Railway-construction-projects/Ostlanken---East-Link-project/}}). HOCHTIEF ViCon, a company that provides services and solutions for digital design and project management, contributed with domain-specific expertise in the activities described in the remainder of this section. 

We want to emphasize that we do not conduct an evaluation study, as such a study would require the assessment of the proposed solution with stakeholders in a natural setting, to improve the solution~\cite{wieringa2014design}. This objective is outside the scope of this paper.

\subsection{Design}
We chose a project from the system engineering domain because our approach requires a domain-specific taxonomy to create trace links. These taxonomies are already established and used in the system engineering domain, and projects in the domain adhere to these taxonomies. Furthermore, a typical use case of trace links is in system verification, where the links help engineers verify that a work product fulfills the stated requirements. Therefore, we collected and analyzed requirements specifications from Eastlink and then verified them against design models that were produced by a sub-contractor.

The study material consists of system requirements, digital 3D models, and a domain-specific taxonomy, SB11. The requirements are written in natural language and consist of two subsets. Project-specific system requirements (240 in total) describe, at varying degrees of detail, the product needs. The generic system requirements consist of 150 documents, each containing dozens to hundreds of requirements. We randomly sampled 27 requirements (10 generic from a document describing how animal fences shall be constructed, and 17 project-specific) for this study. The 3D models contain the design of the railway track, illustrating the location, dimension, and orientation of the physical objects that need to be constructed. The design engineers involved in developing the 3D models have already associated each object with codes from the domain-specific taxonomy, SB11, a classification system from the construction domain that contains more than 2000 classes.

The study was designed and conducted as a series of workshops in the form of online meetings. At least four of the participants listed in Table~\ref{tab:workshops_participants} attended each workshop, where a facilitator (the first author) introduced the task to the participants and asked them for their input.
Hence, to establish trace links between 3D models and requirements, we first had to associate requirements with SB11 codes, and then we investigated how comprehensive and consistent the associations of model objects to SB11 codes were. Furthermore, we verified the traced requirements in the design model.

\paragraph{Requirements classification} We used the functional classification system SB11 to classify the sampled functional requirements. We translated the requirements from Swedish into English using a translation tool. Two of the people who attended the workshops had Swedish language skills and verified the translation. In the workshops, we classified the requirements by reading each requirement and identifying one or more suitable classes from SB11. When we had conflicted opinions, we discussed them until a resolution was achieved. The results of this activity are presented in Section~\ref{sec:associatecodestorequirements}.

\paragraph{Codes verification in design models} The purpose of this activity was to investigate the consistency and reliability of classifying design objects using a classification system (in this case, SB11). Various designers and engineers with different discipline backgrounds, working as sub-contractors, have applied the SB11 codes to a 3D model. If SB11 codes were not applied comprehensively and consistently on objects, tracing those objects to the respective requirements would not be reliable either.
We filtered the objects from the design model by SB11 classes and their sub-classes and verified the object-class association. The verification checked whether the available codes from the SB11 classification system have been comprehensively (all objects are classified) and reliably (all equivalent objects have been classified consistently) applied to the 3D model elements. The results of this activity are presented in Section~\ref{sec:verificationcodesmodels}.

\begin{figure}
  \includegraphics[width=\columnwidth]{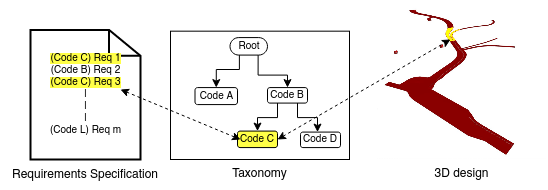}
  \caption{Requirements to Design Objects Tracing}
  \label{fig:req_design_trace}
\end{figure}

\paragraph{Model verification using traced requirements} The purpose of this activity was to verify the conformance of the design models to the specified requirements, as illustrated in Figure~\ref{fig:req_design_trace}. This is one of the main usage scenarios for requirements traceability~\cite{bouillon2013survey}. We selected a class from SB11 and retrieved the requirements and design objects associated with the same class or any sub-classes. We examined each of the retrieved design objects and the related requirements to see if the objects conform to the corresponding requirements. The results of this activity are presented in Section~\ref{sec:modelverif}.

\subsection{Results}

The results of the validation study provide insights into the feasibility of the association of codes to requirements, the verification of codes in the design model, and the usefulness of the created trace links to identify defects in the design model.

\subsubsection{Association of Codes to Requirements}\label{sec:associatecodestorequirements}

\begin{table*}
    \centering
    \footnotesize
    \caption{A sample of requirements that we associated with the classification system SB11}
    \begin{tabular}{llp{0.55\textwidth}l}
        \toprule
        \textbf{Id} & \textbf{Source Document} & \textbf{Requirement} & \textbf{SB11 Codes} \\
        \midrule
        R1 
        & Project specific requirements 
        & The distance to buildings where people are more than temporarily present should be 30 meters in general, taking into account the risk of derailing and other factors. If this distance is not met, the situation should be assessed and choices of departure be justified. 
        & - 
        \\ 
        R2 
        & Project specific requirements 
        & It shall be possible to open the gate to the track from inside the track area by both rescue and evacuation personnel and from outside by rescue personnel. The gate should not be placed closer than 50 m from the tunnel entrance.
        & 18B, 32QG 
        \\
        R3 
        & Generic requirements
        & The opening and closing device must have a width of at least 25m.
        & 32QG
        \\
        R4 & Generic requirements & Fencing should be placed behind the engineering building, seen from the railway, if the property boundary allows this. & 32GDC
        \\
        R5 
        & Project specific requirements
        & Auxiliary power (three-phase, 22 kV) shall be provided along the entire route.
        & 63N
        \\
        R6 & Project specific requirements & Emergency lighting shall be provided in service and access tunnels. & 18B, 63FH\\
        \bottomrule
    \end{tabular}
    \label{tab:RequirementsList}
\end{table*}

We randomly selected 27 natural language requirements originating from two different requirements documents and associated them with codes from SB11. Ten of these requirements were sampled from a document related to wild and game fences while 17 were sampled from the project-specific requirements of Eastlink. Overall, 26 out of the 27 requirements could be associated with at least one code from the SB11 classification. Table~\ref{tab:RequirementsList} shows six representative requirements that illustrate the full set of challenges faced during this activity. In the remainder of this section, we discuss those challenges. 

\paragraph{Challenge 1: Vagueness}
R1 is a safety requirement specifying the minimum distance between two objects. One type of object refers to ``buildings where people are more than just temporarily present'', which is an ambiguous formulation as the timing aspect can be interpreted in different ways and, therefore, results in an unclear set of building types that are affected by this requirement. The second object, the railway track, is implicit and not mentioned in the requirement.  Using only the information that is stated in the requirement, it was not possible to identify any relevant codes from SB11.  
\par
\paragraph{Challenge 2: Compound requirements}
R2 specifies two independent requirements on the same object, i.e., who can open the gate and where it shall be located. This, in turn, means that the requirement contains several objects of interest that need to be associated with the classification system. Compound requirements are not necessarily a problem, can however lead to increased effort in determining all relevant objects in a requirement.
\par
\paragraph{Challenge 3: Context dependency}
R3 refers to an opening and closing device, without however specifying what shall be opened and closed. In this case, the requirement was accompanied by a figure that made clear that the device refers to a gate in a fence. In general, however, for requirements in this group, the textual description alone was not sufficient to understand what objects are relevant. These need to be deduced from the context in which the requirement is mentioned, for example, a particular section in a document or a complete document that covers certain objects of interest. In addition, we could not associate the opening and closing device mentioned in the requirement with any SB11 code. This particular issue is covered under challenge 6: high specificity. 
\par
\paragraph{Challenge 4: Similar classes}
R4 refers to the placement of fencing behind technical buildings. SB11 contains classes related to fences in general, but also in the particular context of road facilities. The correct SB11 class can, therefore, only be deduced by understanding that R4 refers to railway tracks. Furthermore, the decision of association cannot be based solely on the class definitions (they are ambiguous in SB11), but needs to take the hierarchical structure of the taxonomy into consideration. 
\par
\paragraph{Challenge 5: Varying terminology}
R5 specifies that auxiliary power shall be provided along the train track. However, the particular term ``auxiliary power" does not appear in the SB11 taxonomy. It instead contains a code for ``backup, uninterruptible or emergency power" which, one could argue, are either synonyms or more specific instances of auxiliary power. Therefore, associating objects in a requirement with the taxonomy requires domain knowledge on synonyms and specializations of concepts.

\paragraph{Challenge 6: Low specificity}
R6 refers to emergency lighting in service and access tunnels. While emergency lighting exists in SB11 (63FH), there is no code for service or access tunnels, the other relevant objects in the requirement. Hence, we associated these objects with the more generic code for concrete tunnels which is potentially not correct. SB11 is, in this case, not specific enough.

\subsubsection{Verification of Codes in Design Models}\label{sec:verificationcodesmodels}
The purpose of this analysis is to determine whether the subcontractors' application of SB11 to 3D models is comprehensive and reliable. First, to evaluate comprehensiveness, we check if all objects in a 3D model have been associated with an SB11 code. Second, to evaluate reliability, we evaluate whether the same objects have been associated with the same code throughout all 3D models.
We use for this analysis 3D models of two constructions segments which represent a part of the whole facility. For each model, two versions are available: a pre-verification model, as supplied by the sub-contractor to the client, and a post-verification model that contains changes by the sub-contractor after it has been verified by the client. We use letters to refer to the two modeled segments and sub-contractors, i.e. $A$ and $B$, and numerical subscripts to refer to the different versions ($A_1$, $A_2$, $B_1$, $B_2$). Table~\ref{tab:3dmodels} summarizes the properties of the models.

\begin{table}[t]
    \centering
    \footnotesize
    \caption{3D model properties}
    \begin{tabular}{lll}
        \toprule
         \textbf{Model} & \textbf{Sub-contractor} & \textbf{Nr. of objects} \\
        \midrule
        $A_1$ & A & 8435 \\
        $A_2$ & A & 1584 \\
        $B_1$ & B & 306\\
        $B_2$ & B & 380\\
        \bottomrule
    \end{tabular}
    \label{tab:3dmodels}
\end{table}

\begin{figure}
  \centering
  \includegraphics[width=0.8\linewidth]{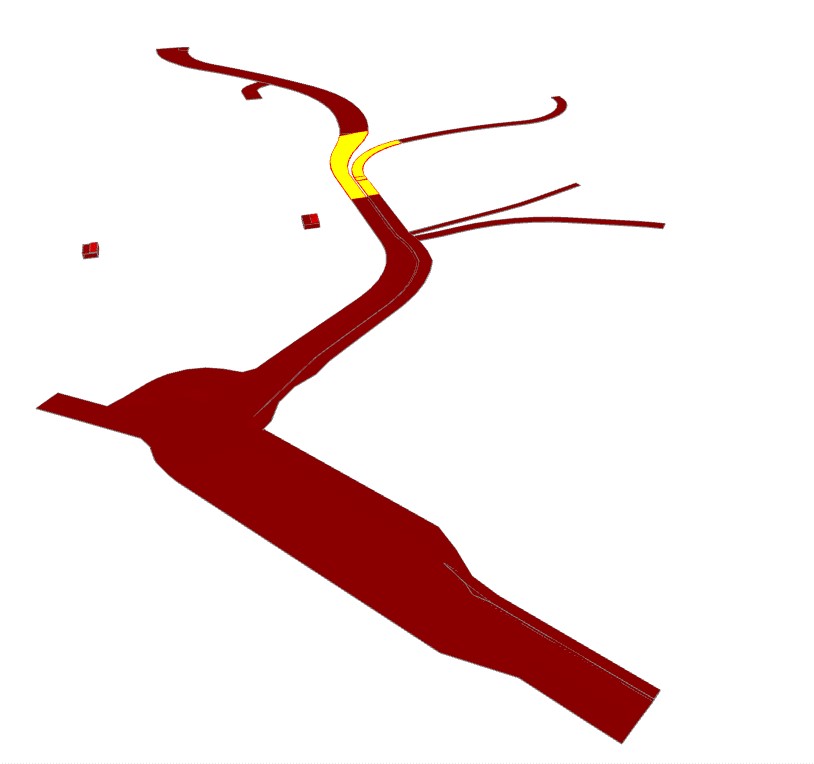}
  \caption{Wrongly classified 3D element.}
  \label{fig:Comprehensiveness1}
\end{figure}

\paragraph{Comprehensiveness}
In all four models, all objects were associated with an SB11 code, indicating that, at least for the studied 3D models, SB11 contained the necessary codes for classification. However, after inspecting the associations in model $A_1$ in more detail, we found that some object classifications were faulty. Some codes contained typos, such as additional or missing characters in the object code. Beyond two syntactic faults, we also found a misclassified object. In Figure~\ref{fig:Comprehensiveness1}, the part of the road highlighted in yellow has been classified as “railway superstructure”, while the correct classification would have been ``road superstructure''. As this fault occurred only once, it is reasonable to assume that it is a human-induced error (e.g., lack of concentration, copy-paste error), similar to the two syntactic faults we found. 

\paragraph{Reliability}
We assessed reliability by checking to what extent equivalent objects, i.e. 3D construction elements of the same type, are consistently classified with the same codes. This consistency can be analyzed from two perspectives: (i) an individual applies codes consistently to equivalent objects over a period of time (intra reliability); (ii) two or more individuals, possibly from different design units or companies, apply codes consistently to equivalent objects (inter reliability). 

Since we had two models available, developed by different subcontractors and at different maturity stages, we were, in principle, able to evaluate both intra and inter reliability. However, as explained next, we encountered several technical difficulties. 

\paragraph{Intra reliability}
To follow an object through different design iterations, the object must be associated with a unique and permanent identifier. While this is the case with the used authoring tool (AutoCAD), the exported format (DWG) we received for analysis did not preserve these identifiers. Hence, analyzing whether the codes associated with an object changed throughout different versions of a model was not possible without additional re-engineering of object identity. We experimented with the following strategies to solve this problem:
\begin{enumerate}
    \item 
    We first attempted to cluster objects in the 3D models based on their geometrical properties. The assumption was that objects belonging to a cluster have similar geometrical properties and hence should be classified similarly. By comparing then a sample of clustered objects from version 1 and version 2 of the two models, we could estimate how consistent the classification is between the two versions. However, we abandoned the idea as we realized that it would be very expensive to manually check the quality of the generated clusters, i.e. that clustered objects are indeed similar or equivalent, given that a model contained thousands of objects.
    \item
    In our second approach, we created an artificial unique ID by means of string concatenation of geometry attributes of 3D elements. The working assumption was that the combination of this multitude of different properties would allow us to trace objects between version 1 and 2 of the model. We used object attributes that describe relative object properties (surface, base, top and lateral areas, as well as volume and center of gravity) and discarded absolute properties (coordinates within the model) to theoretically allow tracing of objects that have been only moved. However, we found that only approximately 30\% of the objects could be traced reliably between the two model versions. This can presumably be explained by the fact that even a minimal change in the object during the transition from one model version to the next led to a change in this very sensitive ID as it consists of many different geometric attributes. Nevertheless, for the objects we \emph{could} trace, we found that the classification did not change.
    \item
    In our third approach, we reviewed and assessed new codes that were introduced in model version 2 and were not used in model version 1. The working assumption was that a new code could indicate a misclassification in version 1 that was corrected in version 2. We found indeed new codes in version 2, but upon inspection of the objects, it turned out that these objects were added in version 2 and did not exist in version 1.
\end{enumerate}

Based on this analysis, we conclude that the application of SB11 codes, i.e. the classification system, has been consistent throughout the two iterations of the model. We need, however, to consider the technical limitations that did not allow us to follow all individual objects between version 1 and version 2 of the model. There might still be inconsistencies in the application of the classification system that we could not detect and that would affect the effectiveness of TTL. Furthermore, we do not know how many designers worked on each model. Our analysis, however, provides some initial evidence that the designers within one subcontractor apply the SB11 codes consistently.

\begin{figure}
  \includegraphics[width=\linewidth]{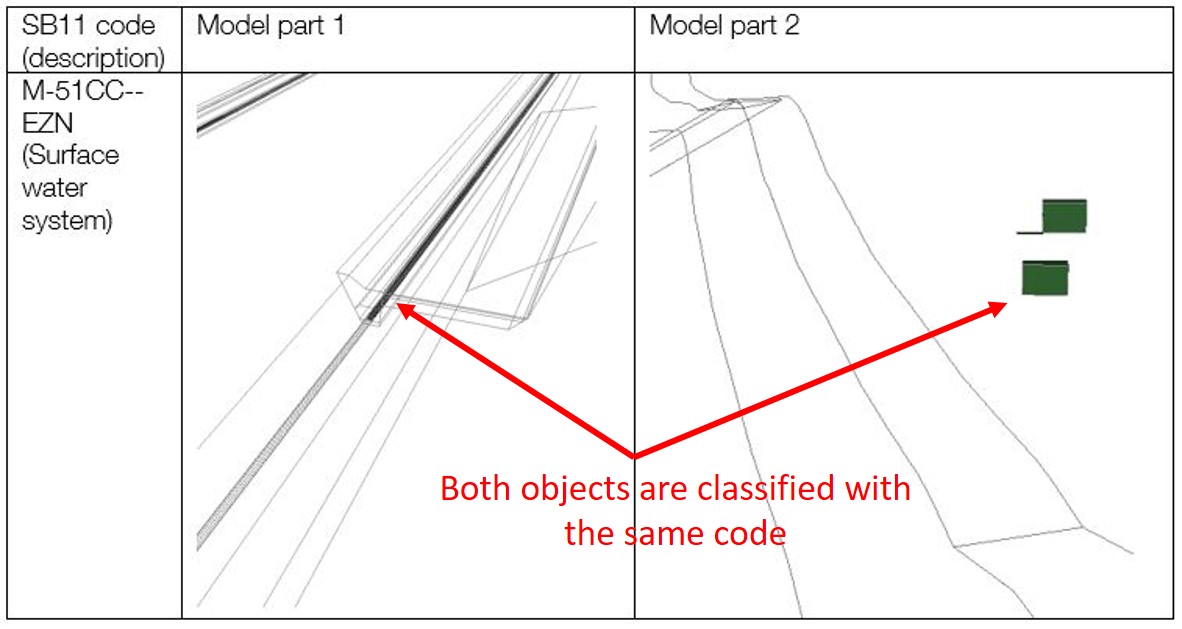}
  \caption{``Surface Water Systems" in two model parts (highlighted in dark color.}
  \label{fig:Reliability1}
\end{figure}

\paragraph{Inter reliability}
To evaluate whether different subcontractors apply the SB11 codes consistently, we compared the classification of objects in version 2 of model A and B. First, we randomly sampled 8 out of the 30 codes. This resulted in 878 (out of 1584) 3D elements from model A. We then visually compared these 878 objects in model A with objects associated with the same code in model B, verifying that the codes are associated with the same type of objects in both models. We found only two objects in which the code SB11 code ``Surface water system'' was applied inconsistently in model A and B. Figure~\ref{fig:Reliability1} shows that two apparent different objects, that are not related, have been classified equally. We speculate again, due to the single occurrence of this fault, that this is due to human error and not an inherent flaw in the classification system or in its process of application. These results provide additional confidence in the feasibility of TTL as it seems that the classification system is applied consistently by different sub-contractors.

\subsubsection{Model verification based on traced requirements}\label{sec:modelverif}
\begin{figure}
  \includegraphics[width=\linewidth]{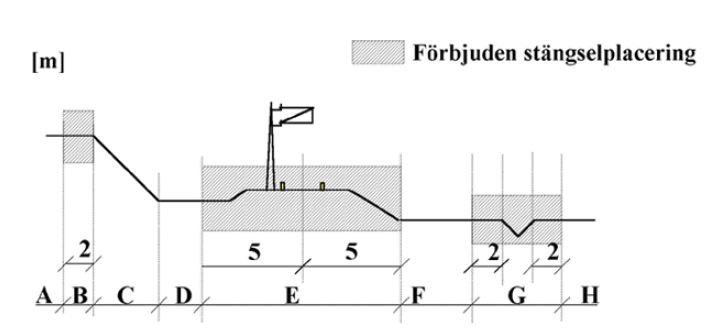}
  \caption{Schematic diagram of the permissible fence position at the surroundings of tracks for requirement K29674. \emph{Note: 'Förbjuden stängselplacering' stands for 'Prohibited fence position'.} The associated requirement text is: ``Fences should be set up in areas A, D, F or H as shown in Figure K5.1, exceptionally may Area C used if fence function and effective maintenance can be ensured."}
  \label{fig:FenceReq}
\end{figure}

We now have the information in place to verify the models against the requirements we traced to SB11. The procedure we followed was:
\begin{enumerate}
    \item Select a requirement from the traced requirements and extract criteria that can be used to verify that the design complies to the requirements. For example, requirement R3 in Table~\ref{tab:RequirementsList} specifies the minimum width of an object.
    \item Filter the model objects by the SB11 code(s) associated with the requirement; in the case of R3, the code is 32QG (gates, gateways).
    \item Verify that the filtered objects fulfill the requirement. 
\end{enumerate}

Out of the 26 requirements we have classified, 11 were associated with codes also present in the models. Ten of these requirements did not reveal any problem with the inspected model objects; a discussion and potential consequences of this observation are provided in Section~\ref{sec:discussion}. For one requirement (K29674), shown in Figure~\ref{fig:FenceReq}, we could identify a potential issue in the inspected models. The requirement defines the areas adjacent to the train tracks where is not allowed to place a fence. Five meters on both sides from the middle of the train tracks (area E) is not allowed to place a fence. In addition, if there is a slope (area C) on the ground from either side of the track, then for at least two meters away of the end of the higher side of the slope (area B), a fence shall not be placed as well. Further than two meters from the end of the slope (area A), on the slope itself (area C) and five meters away from the track until the place where the slope starts (area D), the placement of a fence is allowed.

\begin{figure}
  \includegraphics[width=\linewidth]{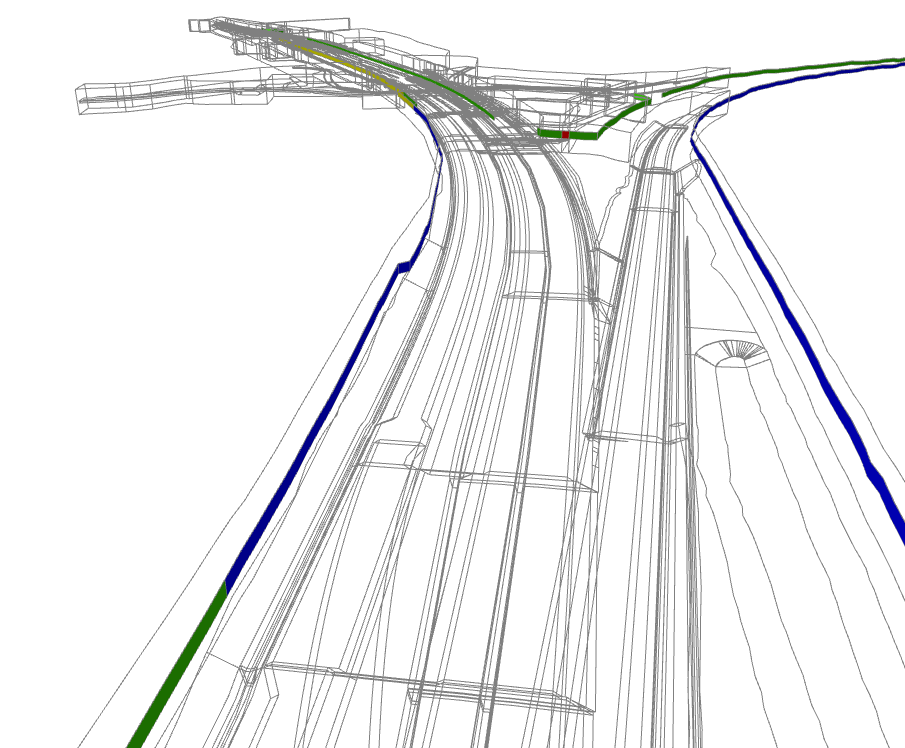}
  \caption{Viewpoint on a BIM Management viewer tool where all elements except from the classified as fences objects  \emph{(32QD--)} have been wired}
  \label{fig:FencesWiredView}
\end{figure}

\begin{figure}
  \includegraphics[width=\linewidth]{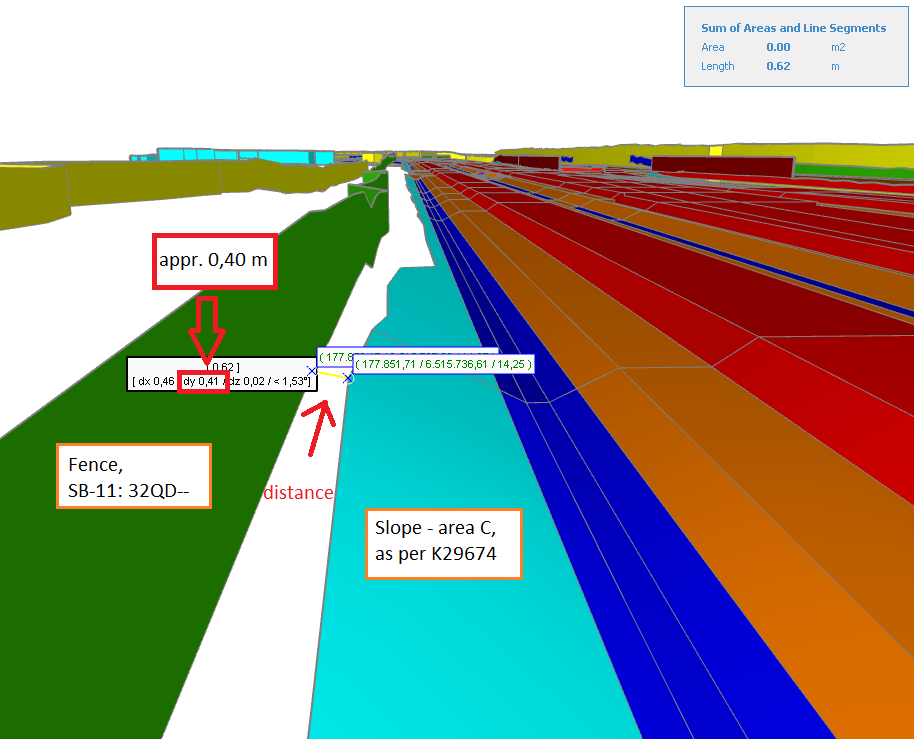}
  \caption{Viewpoint on a BIM Management viewer tool where the fence is placed on the restricted area B as per K29674 requirements definition}
  \label{fig:FenceDistance}
\end{figure}

Filtering the model (model A, version 2) by fence objects (i.e. SB11 code '32QD--' and leaf nodes), highlighted all objects classified as fences (see Figure~\ref{fig:FencesWiredView}). We performed a visual inspection of locations that resembled the schematic diagram in Figure~\ref{fig:FenceReq} and did manual measurements where distances between fence and prohibited areas (B, E, G) seemed to have been violated. We found indeed a few locations where fences were placed within the restricted boundaries of area B. A typical example is depicted in the viewpoint of Figure~\ref{fig:FenceDistance}.

\subsection{Discussion}\label{sec:discussion}
The applicability of TTL in the software domain depends on three dimensions: the requirements, destination artifacts, and the availability of taxonomies.

The majority of \emph{requirements} in software projects are specified in natural language~\cite{kassab2014state, wagner2019status} and both software and system requirements are studied in the same context~\cite{cleland2012software}. In principle, the association of codes, from the taxonomy with the natural language requirements, is similar for systems and software requirements.

\emph{Destination artifacts.}
In software engineering, requirements can be traced to downstream artifacts in the development process, e.g., source code~\cite{mcmillan_combining_2009,hayashi_sentence--code_2010,bavota_role_2013}, architecture~\cite{heindl_case_2005,khan_concern_2009}, design~\cite{ghanam_extreme_2009,buchmann_modelling_2017}, and test cases~\cite{de_lucia_enhancing_2004,lin_poirot_2006,gethers_integrating_2011}. In the case of source code, there is semantic information in the class and method identifiers, compared to 3D models, which contain mainly geometric information. The semantic information and structural information of the source code were considered when creating trace links~\cite{hayashi_sentence--code_2010,bavota_role_2013}. Thus, the automated classification of source code elements using a taxonomy is feasible.

Domain-specific \emph{taxonomies} already exist in the system engineering domain (e.g., SB11), and projects use these taxonomies. However, in software engineering, this depends on the domain of the software product. 
In highly regulated domains, such as medicine~\cite{national_medicalontology} and finance~\cite{cherizola_commodities_2021}, taxonomies or ontologies can be found. In other domains, ontologies can be generated using ontology building techniques~\cite{missikoff_usable_2002,de_nicola_proposal_2005}. 
\subsection{Threats to Validity}
\emph{External validity} concerns the generalization of the results from the validation study in the systems engineering domain to the software engineering domain. First, the TTL approach uses a taxonomy for tracing. Such taxonomies are more likely to be found in highly regulated domains (e.g., SB11 in the infrastructure projects); however, depending on the software product domain, we may need to first build a domain-specific taxonomy. Second, objects in the design models were already associated with codes from the taxonomy by the engineers who produced them. This association, necessary for the acceptance by the customer in the project we studied, seems to be unlikely to be present in the software engineering domain. However, motivation to invest in traceability, and to adopt practical implementations that increase its benefits, is particularly relevant for regulated domains, and product development scenarios that involve sub-contractors and consultants~\cite{mader_motivation_2009}. 

\section{Lessons Learned and Research Roadmap}\label{sec:roadmap}

In this section, we describe a research roadmap for further developing the idea of TTL. We structure the roadmap according to three main components necessary to realize TTL in practice: \emph{reliable classification, domain-specific taxonomies, and tool support and automation}.

\subsection{Reliable Classification}\label{sec:classification}

Implementing the TTL idea can be formulated as a classification problem. Creating TTL requires matching source and destination artifacts with a taxonomy that describes one or more domain dimensions. This matching is a classification of artifacts. 

In general, there are three main types of classification: binary, multi-class, and multi-label. Binary classification refers to assigning one class to an artifact from an output space of two classes. In multi-class classification, we also assign one class to an artifact. However, the output space contains more than two classes. Multi-label is similar to multi-class classification, but an artifact can belong to multiple classes. The classification of artifacts in TTL is a multi-label classification, as multiple classes (labels) from a taxonomy can be assigned to an artifact. As we demonstrated in a previous study~\cite{abdeen_multi_2024}, traditional classifiers are not ready to be used in practice for multi-label classification. Moreover, the taxonomy's characteristics, such as depth, number of nodes, and class description, correlate with the classifier performance.

Thus, a comprehensive and reliable multi-label classification of artifacts is a crucial prerequisite for implementing TTL. If the classification is inconsistent or incomplete, the links may not be reliable, and consequently, the engineers may be unable to use them. Future work needs to investigate what factors can support engineers in reliable multi-label classification.

\subsection{Domains-specific taxonomies}
The study has shown that creating TTL is feasible within the limited context of this study. However, based on sampling and manual evaluation of the classifications made by experts, we identified a few faults and inconsistencies in classifying objects in two models. Thus, it is feasible to use TTL in practice, but the challenges reported in Section~\ref{sec:associatecodestorequirements} need to be addressed.

We need to investigate these challenges further, mainly by looking into the quality of the artifacts to classify and the quality of the taxonomy. Work regarding the latter aspect has already resulted in a compendium of taxonomy quality attributes~\cite{unterkalmsteiner2023compendium}.
 
A main decision when implementing TTL is the choice of domain-specific taxonomy. The taxonomy should be domain-specific, as a general one may not be useful to create trace links on an appropriate level of granularity. These structures may have already been established in well-regulated domains, e.g., finance, business, and infrastructure. Examples of such structures are financial industrial business ontology~\cite{finance_ontology} and the NASA air traffic management ontology~\cite{nasa_air_ontology}. In case of the absence of an established domain-specific structure, one must be created to enable TTL.

Building domain-specific taxonomies and ontologies could be accomplished by using various automated techniques. These techniques create a hierarchical structure from unstructured text by mining domain-specific and general text corpora. Examples of these techniques are OntoGPT~\cite{caufield_structured_2024}, which prompts LLMs to build ontologies; TaxoGen~\cite{zhang_taxogen_2018}, which builds taxonomies by creating clusters from keywords; and TaxoCom~\cite{lee_taxocom_2022}, which produces a taxonomy by extending a partial (seed) taxonomy. Future work needs to investigate the effectiveness of these techniques in building a hierarchical structure that can be used to implement TTL. 

\subsection{Tool Support and Automation}\label{sec:tool}

Automation is crucial to establishing and maintaining TTL. In theory, practitioners can establish TTL manually when creating an artifact. In practice, creating these links could be a daunting task as it requires traversing large taxonomies to find the correct classifications of an artifact. Therefore, it is essential to (semi-)automate trace link creation and maintenance, which is done through the prediction of artifact classification. 

One possible way to achieve automation is by using classifiers that exploit natural language processing (NLP) techniques in combination with large language models such as BERT~\cite{devlin2018bert} to predict classes for textual artifacts. Many software and system development activities produce text. For example, requirements are often written using natural language~\cite{kassab2014state, wagner2019status}, and other artifacts, such as design models, have a textual description for the objects. NLP techniques could be used to process the artifact and taxonomy text to predict classes from the taxonomy, potentially improving the classifier we initiated in earlier work~\cite{unterkalmsteiner_early_2020}. Future work needs to study to what extent machine learning-based classifiers can be used to automate TTL creation and maintenance.

Tool support is also crucial to implement TTL. In general, one barrier that hinders traceability adoption in software projects is the lack of tool support to manage trace links~\cite{ruiz_why_2023}. TTL can be implemented by developing a specific tool for TTL management or integrating it with existing traceability and engineering tools. However, we consider integrating existing tools more beneficial, as it reduces the effort required to adapt a new tool. For example, DesignSpace~\cite{demuth_designspace_2015} is an artifact-agnostic traceability tool that creates direct trace links between artifacts by default but can be customized through the support of a traceability meta-model. Integrating TTL with DesignSpace can bring the benefits of abstraction, structure, and time to the tool, increasing its benefits.

Alternatively, integration with existing engineering tools can be achieved through the development of plugins, such as integration with the requirements management tool DOORS~\footnote{https://www.ibm.com/docs/en/engineering-lifecycle-management-suite/doors/9.7.0?topic=integrating}. The plugin could create trace links between requirements and the taxonomy for inter-requirements tracing. Similar integration with version control tools like Git can be developed to enable tracing of source code to/from other artifacts, e.g., requirements.

\section{Conclusion}\label{sec:conclusion}

In this paper, we explore the idea of taxonomic trace links (TTL), discuss its potential benefits, and provide an initial validation. TTL uses a domain-specific taxonomy to introduce indirection in trace links. TTL has the potential to address important traceability challenges that could hinder implementing reliable trace links, such as the granularity of traces, the lack of a common data structure between artifacts, and unclear responsibility.
Our validation study demonstrates that it is feasible to implement TTL to trace requirements to 3D design models and to verify those requirements in the 3D design model. We also identified challenges that hinder developing a solution that supports engineers in creating TTL. These challenges need to be addressed in future work.

\bibliography{sn-bibliography}

\end{document}